\documentclass{article}
\usepackage{romp40}
\usepackage{amsfonts,amssymb,epsfig}
\usepackage{color}
\newtheorem{thm}{Th\'eor\`eme}[section]
\newtheorem{cor}[thm]{Corollaire}
\newtheorem{lem}[thm]{Lemme}
\newtheorem{pro}[thm]{Proposition}
\newtheorem{dfn}[thm]{D\'efinition}
\newtheorem{rmq}[thm]{Remark}
\newtheorem{expl}[thm]{Exemple}

\title{Coherent states for Landau levels: algebraic and thermodynamical properties}
\author{ Isiaka Aremua \\ 
{\em  Universit\'{e} de Lom\'{e} (UL)} \\
{\em Facult\'{e} Des Sciences  (FDS), D\'{e}partement de Physique,
B.P. 1515 Lom\'{e}, Togo}\\ 
and \\
{\em  International Chair of Mathematical Physics
and Applications} \\
{\em ICMPA-UNESCO Chair, University of Abomey-Calavi,  072 B.P. 50 Cotonou, Republic of Benin}\\  
e-mail:  isiaka.aremua@cipma.uac.bj\\ \\
         Mahouton Norbert Hounkonnou\\
         {\em  International Chair of Mathematical Physics
and Applications} \\
{\em ICMPA-UNESCO Chair, University of Abomey-Calavi,  072 B.P. 50 Cotonou, Republic of Benin}\\  
                      e-mail: norbert.hounkonnou@cipma.uac.bj  \\ \\
                      Ezinvi Balo\"itcha \\
                      {\em  International Chair of Mathematical Physics
and Applications} \\
{\em ICMPA-UNESCO Chair, University of Abomey-Calavi, 072 B.P. 50 Cotonou, Republic of Benin}\\
e-mail:  ezinvi.baloitcha@cipma.uac.bj                    
}

\newcommand{\C}{\mathbb C}

\newcommand{\beq}{\begin{eqnarray}}
\newcommand{\eeq}{\end{eqnarray}}
\newcommand{\bpro}{\begin{pro}}
\newcommand{\epro}{\end{pro}}
\newcommand{\blem}{\begin{lem}}
\newcommand{\elem}{\end{lem}}
\newcommand{\bdfn}{\begin{dfn}}
\newcommand{\edfn}{\end{dfn}}
\newcommand{\bcor}{\begin{cor}}
\newcommand{\ecor}{\end{cor}}
\newcommand{\bthm}{\begin{thm}}
\newcommand{\ethm}{\end{thm}}
\newcommand{\bex}{\begin{expl}}
\newcommand{\eex}{\end{expl}}
\newcommand{\brmq}{\begin{rmq}}
\newcommand{\ermq}{\end{rmq}}
\newcommand{\benum}{\begin{enumerate}}
\newcommand{\eenum}{\end{enumerate}}
\newcommand{\bitem}{\begin{itemize}}
\newcommand{\eitem}{\end{itemize}}

\begin{document}

\maketitle
\begin{abstract}
     This work describes coherent states for a physical system governed by a Hamiltonian operator, 
in two 
dimensional space, of spinless 
charged particles  subject to a perpendicular 
magnetic field ${\bf B},$ coupled with a harmonic potential.
 The underlying $\mathfrak{su}(1,1)$ Lie algebra and 
   Barut-Girardello coherent states are constructed and discussed. Then, the Berezin - Klauder - Toeplitz quantization, also known as
   coherent 
 state (or anti-Wick) quantization,  is   discussed.
The  thermodynamics of such a quantum  gas system is elaborated and analyzed.
\end{abstract}


\section{Introduction}

The system of charged quantum particles interacting with a constant magnetic field is undoubtedly one of the most thoroughly investigated systems 
 in quantum mechanics, mainly inspired by condensed 
matter physics and quantum optics.
A family of coherent states (CS) adapted
to such a system was first proposed in \cite{manko}.  
In  \cite{feldman-kahn}, the behavior of the transverse motion of electrons in an external uniform magnetic field 
${\bf B}$ was considered. A complete set of CS wave packets was constructed.  These states are non-spreading packets 
of minimum uncertainty that follow  a classical motion.  They  are the eigenstates of two non-Hermitian operators 
  that annihilate the 
(zero-) energy  and angular momentum ground states, respectively.   The CS basis was used for the calculation of the partition function. 
Landau diamagnetism and  de Haas-van Alphen oscillations are contained in this setting. 

 Some alternative constructions were proposed in \cite{gazeau-hsiao-jellal} and \cite{schuch-moshinsky}.
In metal and other dense electronic systems, the electrons occupy many Landau levels.  
 Furthermore, the kinetic energy levels of 
electrons in two-dimensional gas correspond to Landau levels. In  \cite{gazeau-novaes}, the generalized 
Gazeau-Klauder CS were extended to systems with more than one degree of freedom. There,  three different types of 
these generalized CS were considered. $\mathfrak{su}(2)$ and $\mathfrak{su}(1,1)$  symmetries of isotopic harmonic oscillator in two-dimensional spinless charged particles 
 in the presence of a constant magnetic field were
investigated.  CS also play an important role in non-equilibrium statistical 
physics.  They describe the evolution towards thermodynamic equilibrium for quantum systems with equidistant energy spectra
\cite{perelomov}.

 CS were also investigated to obtain Landau diamagnetism for a free electron gas \cite{landau2}. 
In \cite{fakhri1} generalized Klauder-Perelomov \cite{perelomov} and Gazeau-Klauder \cite{antoine-gazeau-klauder} CS of Landau 
levels were constructed using two different representations for the Lie algebra $\mathfrak h_{4}$. In \cite{fakhri3}, the Landau levels were 
reorganized into two different hidden symmetries, namely $\mathfrak{su}(2)$ and $\mathfrak{su}(1,1)$. The representation of $\mathfrak{su}(1,1)$ by the Landau levels  then 
led to the construction of the Barut-Girardello  CS (BGCS) \cite{barut-girardello}. 
   In \cite{dehghanietal}, 
 Glauber two-variable 
CS were developed in various representations   using  a unitary displacement operator.
Klauder-Perelomov CS of  $\mathfrak{su}(1,1)$ and
$\mathfrak{su}(2)$ algebras,   for Landau levels, minimizing  
the Heisenberg uncertainty relation, and their statistical properties  were discussed.  
More recently, Bergeron {\it et al} \cite{bergeron}
investigated the consistency of CS quantization,  (also named Berezin-Klauder-Toeplitz or anti-Wick quantization), 
and reached the conclusion 
that the predictions resulting from this type of quantization and those resulting from    canonical quantization are  
compatible on a physical level for
 non-relativistic systems, even though these two quantization techniques are not mathematically equivalent.

This work aims at considering Landau levels 
for  a  Hamiltonian operator describing the motion, 
in two-dimensional space, of spinless charged particles
subject to a perpendicular 
magnetic field ${\bf B}$ coupled with a harmonic potential. 
The underlying $\mathfrak{su}(1,1)$ Lie algebra and 
  BGCS are constructed and discussed. Then, the Berezin - Klauder - Toeplitz quantization  is performed. 
The  thermodynamics of such a quantum  gas system is elaborated and analyzed.

 The paper is organized as follows. In  section $2$,  we start with the study of a spinless charged particles gas  on the $(x,y)$-space 
 in a magnetic 
field ${\bf B}$ with an isotropic harmonic potential.  We use step and  orbit-center coordinate operators. 
 In section $3$, the  $\mathfrak{su}(1,1)$   
representation in the Hamiltonian quantum states is studied. There follows a discussion of the
BGCS of the Lie algebra
 $\mathfrak{su}(1,1).$
    The mean values of {\rm SU}$(1,1)$ group  generators and of the physical system observables,  the  probability density
and the time dependence of the BGCS are calculated.   The quantization of a complex plane 
using these states is investigated in section $4$.  
The Mandel paremeter is studied in section $5.$
In   section $6,$ the main  statistical properties for  
the quantum gas  in  thermodynamic equilibrium  with a reservoir  at temperature $T$ are computed and analyzed.  
Section $7$ is devoted to concluding remarks.  

\section{Hamiltonian operator of an electron in a uniform magnetic field with a harmonic potential}
Consider a system of spinless charged particles  confined to two-dimensional  $(x,y)$-space, with a magnetic field ${\bf B}$ applied along the $z$-direction. 
The eigenstates and eigenvalues of such a system were investigated for the first time by Landau \cite{landau2}. When a harmonic
potential 
is introduced and the Coulomb interactions are neglected, this system is described by the Fock-Darwin Hamiltonian \cite{fock, darwin}
\beq{\label{es0}}
\mathcal H = \frac{1}{2M}\left({\bf p} + \frac{e}{c}{\bf A}\right)^{2} + \frac{M \omega^{2}_{0}}{2}{\bf r}^{2}
\eeq
where $M$ is the particle mass, $e$ the particle charge, ${\bf p}$ is the kinetic momentum and ${\bf A}$ is the vector potential. We study the problem by considering the transverse motion 
of 
the electrons in the $(x,y)$-space.

In the symmetric gauge
\beq{\label{ei18}}
{\bf A} = \left(-\frac{B}{2}y, \frac{B}{2}x \right),
\eeq
the classical Hamiltonian $\mathcal H$ in (\ref{es0}) becomes
\beq{\label{ei19}}
\tilde{\mathcal H}({\bf p},{\bf r}) \equiv \tilde{\mathcal H} = \frac{1}{2M}\left[\left(p_{x} - \frac{eB}{2c}y\right)^{2} + 
\left(p_{y} + \frac{eB}{2c}x\right)^{2}\right] + \frac{M \omega^{2}_{0}}{2}(x^{2} + y^{2}).
\eeq
 Introduce the coordinate  operators  
 $(\hat R_{1}, \hat R_{2}) = (\hat X, \hat Y)$, and momentum operators 
   $(\hat P_{1},  \hat P_{2})   = (\hat P_{x}, \hat P_{y})$,  satisfying the canonical commutation relations
\beq
[\hat R_{i},\hat P_{j}] = i \hbar \delta_{ij}.
\eeq
Let
\beq{\label{ex1}}
\hat Z = \hat X + i \hat Y, \;\;\;  \hat P_{z} = \frac{1}{2}(\hat P_{x} - i \hat P_{y}), \;\;\; 
\hat P_{\bar z} = \frac{1}{2}(\hat P_{x} + i \hat P_{y})
\eeq
and 
consider 
 the  set of energy raising operator $\pi_{+}$ and lowering operator $\pi_{-}$   
defined by \cite{feldman-kahn}
\beq{\label{ei20}}
\pi_{+} = 2\hat P_{\bar{z}} + i\frac{M\Omega}{2}\hat Z,\quad
\pi_{-} = 2 \hat P_{z}- i  \frac{M\Omega}{2}\hat{\bar{Z}}, 
\eeq
such that 
\beq
[\pi_{-},\pi_{+}] = 2M\Omega\hbar.
\eeq
Defining the operators ${\hat \pi_{x}}$ and ${\hat \pi_{y}}$ as:
\beq
 {\hat \pi_{x}}&:=& \hat P_{x} - \frac{M\Omega}{2}\hat Y, \cr
 &&\cr
{\hat \pi_{y}}&:=& \hat P_{y} + \frac{M\Omega}{2} \hat X,
\eeq
we can formulate 
the  angular momentum  raising operator $X_{+}$ and lowering operator $X_{-}$   as follows:
\beq{\label{equ1}}
X_{+} = \left(\hat X - \frac{\hat \pi_{y}}{M\Omega}\right) + i \left(\hat Y + \frac{\hat \pi_{x}}{M \Omega}\right) 
 = \frac{1}{2}\hat Z + \frac{2i }{M \Omega}\hat P_{\bar{z}},
\eeq

\beq{\label{equ2}}
 X_{-} = \left(\hat X - \frac{\hat \pi_{y}}{M\Omega}\right) - i \left(\hat Y + \frac{\hat \pi_{x}}{M \Omega}\right) 
= \frac{1}{2}{\hat{\bar Z}} - \frac{2i }{M \Omega}\hat P_{z}, 
\eeq
with their commutation relation given by
\beq
  [ X_{+},  X_{-}] 
= 2l^{2}.
\eeq
The quantity $l  :=  \sqrt{\frac{\hbar}{M\Omega}}$ is taken as the classical radius of the ground-state's  Landau orbit for the 
frequency $\Omega; $  the motion along the $z$-axis is free,  $\Omega = \sqrt{\omega^{2}_{c} + 4 \omega^{2}_{0}}$ and the cyclotron
 frequency $\omega_{c} = \frac{eB}{Mc}$.

 The eigenvalues  corresponding to (\ref{ei19})  are given by 
\beq{\label{equ11}}
\mathcal E_{n,m} = \hbar \Omega \left(n+\frac{1}{2}\right)
- \frac{\hbar}{2}(\Omega - \omega_{c})m. 
\eeq

 The eigenstates $|n,m\rangle$  are determined by two quantum numbers: $n \in \mathbb N$ associated to the energy,
and $m \in \mathbb Z$ associated to the $z-$ projection of the angular momentum, where the constraint
 $n \geq m$ holds. These eigenvalues are the same as those obtained 
 in \cite{feldman-kahn, dehghanietal}.

 The Hilbert space spanned by the states $|n,m\rangle$ is given by 
 
\beq\label{hilbspace}
\mathfrak H = span\{|n,m\rangle, \, n \geq m, \, n=0, 1, 2, \cdots, +\infty;
m= n, n-1, \cdots, 0, -1,  \cdots, -\infty\}.
\eeq 
 
 
We have the decomposition, for non positive integers $m', $ 

\beq\label{hilbspace00}
\mathfrak H = \left(\bigoplus_{m' < 0}\mathfrak H'_{m'}\right)   \bigoplus \left(\bigoplus_{m=0}^{+\infty}\mathfrak H_{m}\right)
\eeq

such that the identity operator $I_{\mathfrak H}$ on $\mathfrak H$ writes   as follows:

\beq\label{hilbspace01}
I_{\mathfrak H} = \left(\bigoplus_{m' < 0}I_{\mathfrak H'_{m'}}\right)   
\bigoplus \left(\bigoplus_{m=0}^{+\infty}I_{\mathfrak H_{m}}\right) 
\eeq
 
 with 

\beq
I_{\mathfrak H'_{m'}} = \sum_{n=0}^{+\infty}|n, m'\rangle \langle n, m'|, \qquad  
I_{\mathfrak H_{m}} = \sum_{n=m}^{+\infty}|n,m\rangle \langle n,m|.
\eeq
 

\section{Representation of $\mathfrak{su}(1,1)$ algebra of the quantum Hamiltonian  states}
\subsection{ Hilbert space representation}
Rewrite now the step and orbit-center coordinate operators, denoted by  $\hat{\pi}_{\pm}$ and $\hat{ X}_{\pm}, $  
with the help of dimensionless 
variables as follows:
\beq
\hat \pi_{+} = l\sqrt{2}\left[\frac{i}{4l^{2}}\hat Z + \frac{1}{\hbar}\hat P_{\bar z}\right], \quad \hat \pi_{-} = 
l\sqrt{2}\left[-\frac{i}{4l^{2}}\hat{\bar Z} + \frac{1}{\hbar}\hat P_{z}\right], 
\eeq
\beq
\hat{ X}_{+} = \frac{\sqrt{2}}{l}\left[\frac{1}{4}\hat Z + \frac{i l^{2}}{\hbar}\hat P_{\bar z}\right], \quad \hat{ X}_{-} 
= \frac{\sqrt{2}}{l}\left[\frac{1}{4}\hat{\bar Z} - \frac{i l^{2}}{\hbar}\hat P_{z}\right].
\eeq
  These operators satisfy the following canonical commutation relations
\beq{\label{eq007}}
[\hat{\pi}_{-},\hat{\pi}_{+}] = \mathbb I,  \qquad [\hat{ X}_{+}, \hat{ X}_{-}]= \mathbb I.
\eeq
The application of ${ X}_{+}$ on the highest weight vector 
$|n,n \rangle$ yields $0$ \cite{feldman-kahn}. 

Consider the  Lie algebra $\mathfrak{su}(1,1)$ \cite{gilmore} corresponding to the {\rm SU}$(1,1)$ group, 
spanned by the three group generators $\{\mathcal K_{1}, \mathcal K_{2}, \mathcal K_{3}\}$ such that
\beq
[\mathcal K_{1}, \mathcal K_{2}] = -i \mathcal K_{3}, \quad [\mathcal K_{2}, \mathcal K_{3}] = i \mathcal K_{1}, 
\quad [\mathcal K_{3}, \mathcal K_{1}] = i \mathcal K_{2}.
\eeq



As matter of convenience, let us define 
the raising and lowering operators $\mathcal K_{\pm} = \mathcal K_{1} \pm i \mathcal K_{2}$
as the  following second-order differential operators:

\beq
\mathcal K_{+}: = \hat{\pi}_{+} \hat{X}_{-}, \quad \mathcal K_{-}: = \hat{\pi}_{-}\hat{ X}_{+}, 
\eeq
and the operator

\beq 
\mathcal K_{3} := \frac{1}{2}(\hat{\pi}_{+}\hat{\pi}_{-} + \hat{ X}_{+}\hat{ X}_{-}).
\eeq
They satisfy  the commutation relations of 
the Lie algebra $\mathfrak{su}(1,1)$:
\beq{\label{eq23}}
[\mathcal K_{+}, \mathcal K_{-}] = -2 \mathcal K_{3}, \qquad [\mathcal K_{3}, \mathcal K_{\pm}] = \pm \mathcal K_{\pm}. 
\eeq
The operators $\hat{\pi}_{+}$ and $\hat{\pi}_{-}$ act as follows:
\beq{\label{eq180}}
\hat{\pi}_{+}|n-1,m-1\rangle = \sqrt{n}|n,m\rangle,
\quad
\hat{\pi}_{-}|n,m\rangle = \sqrt{n}|n-1,m-1\rangle,
\eeq
\beq{\label{eq1090}}
\hat{ X}_{+}|n,m\rangle = \sqrt{(n-m)}|n,m+1\rangle, \quad
\hat{ X}_{-}|n,m\rangle = \sqrt{(n-m+1)}|n,m-1\rangle.
\eeq




The state   $|m,m\rangle$ is annihilated by the 
lowering operator $\mathcal K_{-}$, i.e.
\beq
\mathcal K_{-}|m,m\rangle = 0.
\eeq
There results the following  representation of the Lie algebra $\mathfrak{su}(1,1)$ in  $\mathfrak H$ given in (\ref{hilbspace}):
\beq{\label{rep2}}
\mathcal K_{+}|n-1,m\rangle &=&  \sqrt{n(n-m)}|n,m\rangle \crcr
\mathcal K_{-}|n,m\rangle &=&  \sqrt{n(n-m)}|n-1,m\rangle \crcr
\mathcal K_{3}|n,m\rangle  &=& \frac{1}{2}(2n-m+1)|n,m\rangle.
\eeq

We deduce the expression of an  arbitrary state $|n,m\rangle$ using the first equation of 
(\ref{rep2}) as follows:
\beq
|n,m\rangle = \sqrt{\frac{\Gamma(m+1)}{\Gamma(n-m+1)\Gamma(n+1)}} \,\mathcal K^{n-m}_{+}|m,m\rangle.
\eeq

\subsection{CS for the $\mathfrak{su}(1,1)$ algebra}

In this paragraph, we first construct the  $\mathfrak{su}(1,1)$ algebra  CS for the quantum system on the 
 Hilbert subspace $\mathfrak H_{m}:=span\{|n,m\rangle\}_{n \geq m, m\geq 0} $ and after on the subspace 
 $\mathfrak H'_{m'}:=span\{|n, m'\rangle\}_{n \geq m', m' <  0}. $
Then we discuss their    reproducing kernel 
 and  analyticity. 
 

The BGCS 
 in the Hilbert subspace $\mathfrak H_{m}$ can be defined as eigenstates 
of the lowering generator $\mathcal K_{-}$ of the Lie algebra $\mathfrak{su}(1,1)$ \cite{barut-girardello}, i.e.
\beq{\label{gen00}}
\mathcal K_{-}|z\rangle_{m} = z|z\rangle_{m}
\eeq
where $z$ is an arbitrary complex variable of the form $z = \rho e^{i \phi}, 0 \leq \rho < \infty, 0 \leq \phi < 2\pi.$ 
The eigenstates $|z\rangle_{m}$ can be represented as the superposition of the complete orthonormal basis $|n,m\rangle$ of 
$\mathfrak H_{m}$ as follows:
\beq{\label{gen01}}
|z\rangle_{m} = \sum_{n=m}^{+\infty}\langle n,m|z\rangle_{m}  \,  |n,m\rangle.
\eeq

Using (\ref{gen00}), (\ref{gen01}), the third equation in 
(\ref{rep2}) and the orthonormality relation
\beq{\label{gen02}}
\sum_{n=m}^{+\infty}|n,m\rangle\langle n, m| = I_{\mathfrak H_{m}},
\eeq
we get
\beq{\label{gen03}}
\langle n,m|z\rangle_{m} = \frac{z}{\sqrt{n(n-m)}}\langle n-1,m|z\rangle_{m}.
\eeq
This relation  can be recursively transformed into
\beq
\langle n,m|z\rangle_{m} = \sqrt{\frac{\Gamma(m+1)}{\Gamma(n-m+1)\Gamma(n+1)}} \, z^{n-m} \langle m,m|z\rangle_{m}.
\eeq
Then  the state $|z\rangle_{m}$ in the Hilbert subspace 
$\mathfrak H_{m}$ can be rewritten as:
\beq{\label{gen04}}
|z\rangle_{m} = |\langle m,m|z\rangle_{m}|\sum_{n=m}^{+\infty}z^{n-m}
\sqrt{\frac{\Gamma(m+1)}{\Gamma(n-m+1)\Gamma(n+1)}} |n,m\rangle.
\eeq
The normalization factor
$|\langle m,m|z\rangle_{m}|$ can be obtained by
 normalizing to unity the states $|z\rangle_{m}$ and using the relation  \cite{magnus}
\beq{\label{rec01}}
\sum_{n=0}^{+\infty}\frac{x^{n}}{n !\Gamma(n+\nu+1)} = \frac{1}{x^{\nu}}I_{\nu}(2x),
\eeq
where $I_{\nu}(2x)$ is the modified Bessel function of order $\nu.$ It follows
\beq
|\langle m,m|z\rangle_{m}| = \sqrt{\frac{|z|^{m}}{I_{m}(2|z|) \,\Gamma(m+1)}}.
\eeq
Finally, the eigenstates $|z\rangle_{m}$ become
\beq{\label{gen05}}
|z\rangle_{m} = \frac{|z|^{m/2}}{\sqrt{I_{m}(2|z|)}}
\sum_{n=m}^{+\infty}\frac{z^{n-m}}{\sqrt{\Gamma(n-m+1)\Gamma(n+1)}}|n,m\rangle
\eeq
where $I_{m}(2|z|)$ is the modified Bessel function of the first kind given by \cite{magnus}:
\beq{\label{gen06}}
I_{m}(2|z|) = \sum_{n=0}^{+\infty}\frac{|z|^{2n+m}}{\Gamma(n+1)\Gamma(n+m+1)}.
\eeq
These states satisfy the following resolution of the identity \cite{barut-girardello}:
\beq{\label{gen07}}
\int_{\mathbb C}|z\rangle_{m} \, _{m}\langle z|d\varrho(z) = I_{\mathfrak H_{m}}
\eeq
 on  $\mathfrak H_{m};$   $d\varrho(z)$ is an appropriate measure. Indeed, performing 
the integrals  over the whole complex plane, where $z = \rho e^{i \phi},  \rho \in [0,\infty),  \phi \in [0,2\pi),$ and 
taking
\beq{\label{gen008}}
d\varrho(z) = \frac{2}{\pi}I_{m}(2|z|)K_{m}(2|z|)d^{2}z, \;\;\; d^{2}z = d(Re \, z) d(Im \, z),
\eeq
  the relation (\ref{gen02}) defined on $\mathfrak H_{m}$ leads to  the resolution of the identity (\ref{gen07})  
by using the following  relation:
\beq{\label{gen08}}
4 \int_{0}^{\infty}\rho^{2n - m + 1}K_{m}(2\rho)d\rho = \Gamma(n-m+1)\Gamma(n+1)
\eeq
deduced from  the  integral \cite{watson}:
\beq{\label{stat007}}
\int_{0}^{\infty}dx \, x^{\mu}K_{\nu}(ax) = 2^{\mu - 1}a^{-\mu-1}\Gamma\left(\frac{1+\mu + \nu}{2}\right)\Gamma\left(\frac{1+\mu 
- \nu}{2}\right), \crcr
 \ [Re(\mu + 1 \pm \nu) > 0, Re \,a > 0 ].
\eeq
The quantity $K_{m}(2\rho),$ given by
\beq
K_{m}(2\rho) = \frac{\pi}{2}\, \frac{I_{-m}(2\rho) - I_{m}(2\rho)}{\sin{(m\pi)}},
\eeq
is the modified Bessel function of the second kind \cite{magnus}.

The states $|z\rangle_{m}$ form an overcomplete basis for any allowed value of $m$. To prove this, use (\ref{gen04}),  
and set $z = \rho e^{i \phi}$, where $\rho \in [0,\infty),  \phi \in [0,2\pi).$  It follows that
\beq\label{eqresolv}
&&\int_{\mathbb C}|z\rangle_{m}\, _{m}\langle z|d\varrho(z)\crcr
&&= 4 \sum_{n=m}^{+\infty} \int_{0}^{\infty} \frac{K_{m}(2\rho)}{\Gamma(n-m+1)\Gamma(n+1)} \rho^{2n-m+1} 
d\rho|n,m\rangle \langle n,m|.
\eeq
Then, using (\ref{gen08}), we get on the Hilbert subspace $\mathfrak H_{m}$:
\beq\label{resolv}
\int_{\mathbb C}|z\rangle_{m}\, _{m}\langle z|d\varrho(z) = \sum_{n=m}^{+\infty}|n,m\rangle \langle n,m| 
= I_{\mathfrak H_{m}}.
\eeq



We have for $n\geq 0$ and $m' <  0$

\beq
 |n,m'\rangle = 
\sqrt{\frac{\Gamma(-m'+1)}{\Gamma(n-m'+1)\Gamma(n+1)}}\, \, \mathcal K^n_{+}|0, m'\rangle.
\eeq

\bpro
The components of the BGCS $|z\rangle_{m'}  
$  such that $\mathcal K_{-}|z\rangle_{m'} =  z |z\rangle_{m'} $ expanded as 

\beq
|z\rangle_{m'}  = 
\sum_{n=0}^{+\infty}\langle n,m'|z\rangle_{m'}|n, m'\rangle 
\eeq

are given by

\beq\label{orthonorm}
\langle n, m'|z\rangle_{m'} = \sqrt{\frac{\Gamma(-m'+1)}{\Gamma(n-m'+1)\Gamma(n+1)}}\,z^n 
\langle 0, m'|z\rangle_{m'} 
\eeq

where 

\beq
|\langle 0,m'|z\rangle_{m'}| = \sqrt{\frac{|z|^{-m'}}{I_{-m'}(2|z|) \,\Gamma(-m'+1)}}. 
\eeq

\epro

{\bf Proof.}  It uses the third equation of (\ref{rep2}), the orthonormality relation and 
equation (\ref{orthonorm}).

$\hfill{\blacksquare}$

The BGCS
 for $m' <  0, $
 denoted by $|z\rangle_{m'}  
$, corresponding 
to the constraint ${\bf B}. {\bf L} \leq 0$ with ${\bf L}$ the 
angular momentum, 
are given  in the   subspace  $\mathfrak H'_{m'}=span\{|n, m'\rangle\}_{n \geq m', m' <  0}$ through the relation

\beq\label{generat06} 
|z\rangle_{m'} = \frac{|z|^{-m'/2}}{\sqrt{I_{-m'}(2|z|)}}
\sum_{n=0}^{+\infty}\frac{z^{n}}{\sqrt{\Gamma(n-m'+1)\Gamma(n+1)}}|n, m'\rangle, 
\quad m' <  0.
\eeq

The resolution of the identity on the  Hilbert subspace $\mathfrak H'_{m'}=span\{|n, m'\rangle\}_{n \geq m', m' <  0}$ 
is given by 
\beq\label{resolv00}
\int_{\C}|z\rangle_{m'}\, _{m'}{\langle z}|d\tilde\varrho(z) 
= \sum_{n=0}^{+\infty}|n, m'\rangle \langle n, m'|= I_{\mathfrak H'_{m'}} 
\eeq

where the modified Bessel function $ K_{m'}(2\rho) = K_{-m'}(2\rho), m' < 0$  and $d\tilde\varrho(z)$ satisfy
an analog equality as (\ref{eqresolv}).

%



The resolution of the identity on the entire Hilbert space $\mathfrak H$ is deduced from (\ref{hilbspace01}), (\ref{resolv}) and 
(\ref{resolv00}) as

\beq
I_{\mathfrak H} = \sum_{m' < 0}\int_{\C}|z\rangle_{m'}\, _{m'}{\langle z}|d\tilde\varrho(z) \bigoplus
\sum_{m=0}^{+\infty}\int_{\mathbb C}|z\rangle_{m}\, _{m}\langle z|d\varrho(z).
\eeq


Besides, the BGCS  (\ref{gen05}) are not orthogonal, i.e.  given two vectors $|z\rangle_{m}$ and 
$|z'\rangle_{m'},$ ($m\neq m'$), on the Hilbert space 
$\mathfrak H_{m}, $ 
 their inner product is not null:
\beq\label{exprs00}
_{m}\langle z'|z\rangle_{m} = \left(\frac{|z'z|}{{\bar{z'}}z}\right)^{m/2}
\frac{I_{m}\left(2\sqrt{{\bar{z'}}z}\right)}{\sqrt{I_{m}(2|z'|)I_{m}(2|z|)}}\neq 0.
\eeq
The overcompleteness of the BGCS $|z\rangle_{m}$ on  $\mathfrak H_{m}$  suggests to  discuss their  relation  with the 
reproducing   kernels   \cite{ali-antoine-gazeau}. 
 Define the quantity $\mathcal K(z, z') := \,_{m}\langle z'|z\rangle_{m}$ on  $\mathfrak H_{m}$. Using  the facts

\beq
\overline{_{m}\langle z'|z\rangle_{m}} &=& \left(\frac{|zz'|}{ \bar z z'}\right)^{m/2}
\frac{1}{\sqrt{I_{m}(2|z|)I_{m}(2|z'|)}}\crcr
&& \times \sum_{n=m}^{+\infty}(\bar z z')^{(2n-m)/2}
\frac{1}{\Gamma(n-m+1)\Gamma(n+1)},
\eeq

\beq
|\bar z z'|^{m/2} = \left(\frac{|z'z|}{{\bar{z}}z'}\right)^{m/2}({\bar{z}}z')^{m/2}
\eeq

and setting $n=m+\nu$, it follows that  

\beq\label{exprs01}
\overline{_{m}\langle z'|z\rangle_{m}} = \left(\frac{|zz'|}{ \bar z z'}\right)^{m/2}
\frac{I_{m}\left(2\sqrt{\bar z z'}\right)}{\sqrt{I_{m}(2|z'|)I_{m}(2|z|)}} =:  \mathcal K(z', z).
\eeq
 $\mathcal K(z, z')$ is well a  reproducing kernel. Indeed, 
 
\bpro
The following properties 

\bitem

\item [(i)] hermiticity
\beq
\mathcal K(z, z') = \overline{ \mathcal K(z', z)}, 
\eeq

\item  [(ii)]  positivity

\beq
\mathcal K(z,z) > 0,
\eeq

\item  [(iii)] idempotence
\beq\label{exprs02}
\int_{\mathbb C}d\varrho(z'') \mathcal K(z, z'') \mathcal K(z'', z') = \mathcal K(z, z') 
\eeq

\eitem

are satisfied by the function $\mathcal K$ on $\mathfrak H_{m}$. 
\epro

{\bf Proof.} (i) and (ii) result  from  (\ref{exprs01}) and (\ref{exprs00}), and
(\ref{exprs02}) from direct computation.

$\hfill{\blacksquare}$

From (\ref{resolv}), for any $|\Psi\rangle \in \mathfrak H_{m}$, we have 

\beq
|\Psi \rangle  =   \int_{\mathbb C}d\varrho(z) \Psi(z)  |z\rangle_{m} 
\eeq
where $\Psi(z) := \,  _{m}\langle  z|\Psi\rangle $. 
The following reproducing property 

\beq
\Psi(z) = \int_{\mathbb C}d\varrho(z') \mathcal K(z,z') \Psi(z')
\eeq
is also verified.

The Hilbert space    $\mathfrak H_{m}$ 
can be represented as the Hilbert space of analytic functions in the variable $z$.  
Given a normalized state $|\Phi\rangle = \sum_{k=m}^{+\infty}C_k|k,m\rangle, C_k \in \mathbb C$ on $\mathfrak H_{m}$, we obtain 

\beq
_{m}\langle \bar z|\Phi\rangle =  \frac{|z|^{m/2}}{\sqrt{I_{m}(2|z|)}}
\sum_{n=m}^{+\infty}\frac{ C_n \, z^{n-m}}{\sqrt{\Gamma(n-m+1)\Gamma(n+1)}} 
\eeq 
such that the entire functions 
\beq
f(z;m) = \frac{\sqrt{I_{m}(2|z|)}}{|z|^{m/2}}\, _{m}\langle \bar z|\Phi\rangle  = 
\sum_{n=m}^{+\infty}\frac{ C_n \, z^{n-m}}{\sqrt{\Gamma(n-m+1)\Gamma(n+1)}} 
\eeq
are analytic over the whole $z$ plane.  Then,
from (\ref{resolv}), we can write 

\beq
|\Phi\rangle  = \int_{\mathbb C}d\varrho(z)\, \frac{|z|^{m/2}}{\sqrt{I_{m}(2|z|)}}\,  f(\bar z;m)|z\rangle_{m} 
\eeq
and express the scalar product of two states $|\Phi_1\rangle$ and $|\Phi_2\rangle$ given on $\mathfrak H_{m}$ by
the formula
\beq
\langle \Phi_1|\Phi_2 \rangle = \int_{\mathbb C}d\varrho(z)\, \frac{|z|^{m }}{ I_{m}(2|z|) }\,  \overline{f_1(\bar z;m)}f_2(\bar z;m).  
\eeq

\subsection{Mean values}\label{paragraph}


 The $\mathfrak{su}(1,1)$ CS  can be used
 in different physical applications to calculate the expectation (mean) values of any significant 
 physical observable $\mathcal O$ which characterizes the quantum system
embedded in the harmonic potential. 

Indeed, using equation (\ref{gen05}), the mean value of a physical observable $\mathcal O$ in
the  BGCS $|z\rangle_{m}$ is obtained as:
\beq{\label{gen09}}
_{m}\langle z|\mathcal O|z\rangle_{m} \equiv  \langle \mathcal O\rangle_{z,m} &=& \frac{|z|^{m}}{I_{m}(2|z|)}
\sum_{n,k=m}^{+\infty} z^{n-m}{\bar z}^{k-m}
\sqrt{\frac{1}{\Gamma(n-m+1)\Gamma(n+1)}} \crcr
&&\times \sqrt{\frac{1}{\Gamma(k-m+1)\Gamma(k+1)}} \langle k,m|\mathcal O|n,m\rangle.
\eeq
Setting $n=m+\nu$, $k=m+\upsilon$, respectively, (\ref{gen09}) can be rewritten as follows:
\beq{\label{gen10}}
\langle \mathcal O\rangle_{z,m} &=& \frac{|z|^{m}}{I_{m}(2|z|)}\sum_{\nu,\upsilon=0}^{+\infty}
\frac{{\bar{z}}^{\upsilon}z^{\nu}}
{\sqrt{\Gamma(\nu+1)\Gamma(\nu+m+1)\Gamma(\upsilon+1)\Gamma(\upsilon+m+1)}}\crcr
&& \times \langle \upsilon,m|\mathcal O|\nu,m\rangle.
\eeq
The computation of the mean values of the $\mathcal K_{i},  (i=-,+), $ using (\ref{rep2}) and (\ref{gen10}), gives
\beq{\label{gen11}}
\langle \mathcal K_{-}\rangle_{z,m} &=& z\frac{1}{I_{m}(2|z|)}\sum_{\nu=0}^{+\infty}
\frac{|z|^{2\nu + m}}
{\Gamma(\nu+1)\Gamma(\nu+m+1)} = z,
\eeq
and
\beq{\label{gen12}}
\langle \mathcal K_{+}\rangle_{z,m} &=& {\bar z}\frac{1}{I_{m}(2|z|)}\sum_{\nu=0}^{+\infty}
\frac{|z|^{2\nu + m}}
{\Gamma(\nu+1)\Gamma(\nu+m+1)} = {\bar z}.
\eeq
We can conclude, therefore, that  $\langle \mathcal K_{-}\rangle_{z,m}$ and $
\langle \mathcal K_{+}\rangle_{z,m}$ are mutually conjugated.

Since the generators $\mathcal K_{+}, \mathcal K_{-}$ are given by 
$\mathcal K_{\pm} = \mathcal K_{1} \pm i \mathcal K_{2}$, 
we obtain from (\ref{gen11}) and (\ref{gen12}) the following expressions:
\beq
\langle \mathcal K_{1}\rangle_{z,m} = \frac{1}{2}\langle \mathcal K_{-}+\mathcal K_{+}\rangle_{z,m} 
= \frac{1}{2}(z+{\bar z}) = Re \, z,
\eeq
\beq
\langle \mathcal K_{2}\rangle_{z,m} = \frac{i}{2}\langle \mathcal K_{-}-\mathcal K_{+}\rangle_{z,m} 
= \frac{i}{2}(z-{\bar z}) = -Im \, z.
\eeq

In order to compute the mean values of the generator $\mathcal K_{3}$ and its square $\mathcal K^{2}_{3}$, 
it is useful to evaluate  the sum $S_{n}$, with $n=0,1,2, \dots$, given by (see Appendix in \cite{popov}):
\beq
S_{n} = \sum_{\nu = 0}^{+\infty}\frac{(x^{2})^{\nu}}{\Gamma(\nu + 1)\Gamma(\nu + n + 1)}\,\nu^{n}.
\eeq
Then, from (\ref{rep2}), we obtain
\beq{\label{gen14}}
\langle \mathcal K_{3}\rangle_{z,m}  =  
|z| \frac{I_{m+1}(2|z|)}{I_{m}(2|z|)} + \frac{m+1}{2}
\eeq
and
\beq{\label{gen16}}
\langle \mathcal K^{2}_{3}\rangle_{z,m} 
= |z|^{2}\frac{I_{m+2}(2|z|)}{I_{m}(2|z|)} + (m+2)|z| \frac{I_{m+1}(2|z|)}{I_{m}(2|z|)} + \left(\frac{m+1}{2}\right)^{2}.
\eeq
Besides, the number operator $N$ 
 diagonalizing the basis vectors $\{|\nu, m\rangle, \nu \geq 0 \}$ for the number states, 
\beq\label{numb}
N|\nu, m\rangle = \nu |\nu, m\rangle,
\eeq
is used to compute the photon number distribution  as follows:
\beq{\label{ph}}
|\langle \nu,m|z\rangle_{m}|^{2} = \frac{|z|^{2\nu + m}}{I_{m}(2|z|) \,\Gamma(\nu+1)\Gamma(\nu + m+1)}.
\eeq
Exploiting  (\ref{gen14}) and (\ref{gen16}), we are able to compute explicitly the mean values of the number operator and its second power 
to obtain:
\beq{\label{numb00}}
\langle N \rangle_{z,m} &=& \left\langle \mathcal K_{3}  - \frac{m+1}{2} \right\rangle_{z,m} = |z| \frac{I_{m+1}(2|z|)}{I_{m}(2|z|)},\cr
\langle N^{2} \rangle_{z,m}  &=&  \left\langle \mathcal K^{2}_{3}  - (m+1)\mathcal K_{3} +  
\left(\frac{m+1}{2}\right)^{2}\right\rangle_{z,m}  =
 |z|^{2}\frac{I_{m+2}(2|z|)}{I_{m}(2|z|)} + |z| \frac{I_{m+1}(2|z|)}{I_{m}(2|z|)}.\nonumber
 \\
\eeq
Then, it becomes straightforward to extend the calculation to the intensity  correlation,   highlighting the bunching and antibunching 
effects of the quantum states, defined as in \cite{ brif-aryeh}:
\beq
g^{(2)}_{z,m} = \frac{\langle N^{2} \rangle_{z,m} - \langle N \rangle_{z,m}}{\langle N \rangle^{2}_{z,m}} 
= \frac{I_{m}(2|z|)I_{m+2}(2|z|)}{[I_{m+1}(2|z|)]^{2}}.
\eeq
More specifically, for two interesting limiting cases of the $|z|$ variable, i.e.  for $|z| \ll 1$ and $|z| \gg 1$, using the approximations 
for the Bessel modified function $I_{m}(x)$ \cite{gradshtein}
\beq\label{approxim}
I_{m}(x) \simeq \frac{1}{\Gamma(m+1)}\left(\frac{x}{2}\right)^{m},
\quad
I_{m}(x) = \frac{e^{x}}{\sqrt{2\pi x}}\left[1+ O \left(\frac{1}{x}\right)\right],
\eeq
we get for the intensity correlation function, with
\beq\label{approx}
I_{m}(2|z|) \simeq \frac{|z|^{m}}{\Gamma(m+1)}, \; I_{m+1}(2|z|) \simeq \frac{|z|^{m+1}}{\Gamma(m+2)}, \; 
I_{m+2}(2|z|) \simeq \frac{|z|^{m+2}}{\Gamma(m+3)},
\eeq
the following expressions:
\beq
g^{(2)}_{z,m} \simeq \frac{m+1}{m+2},
\eeq
\beq
g^{(2)}_{z,m} \simeq 1,
\eeq
respectively.
Thus, for small values of $|z|$, the intensity correlation function is smaller than unity, for all values of $m$. 
The corresponding  BGCS $|z\rangle_{m}$ exhibit sub-Poissonian statistics behaviour
  (with antibunching effect, i.e.  $g^{(2)}_{z,m}  < 1$), while for large $|z|$, 
these states tend to have Poissonian statistics  (case of the standard CS,  $g^{(2)}_{z,m} \simeq 1$). In addition, 
the photon-number distribution (\ref{ph}) is sub-Poissonian \cite{ brif-aryeh}.

\subsection{Observables of the physical system mean values}

The operators $\pi_+, \pi_-$ and $X_+, X_-$ of the system, provided in (\ref{ei20}), (\ref{equ1}) and (\ref{equ2}), 
are used to express the coordinate operators $\hat X, \hat Y$ and 
momentum operators $\hat P_x, \hat P_y$ as follows:

\beq
\hat X   =  \frac{1}{2}\left[ X_{+} +  X_{-}\right] 
 - \frac{i}{2M\Omega}\left[ \pi_{+} -   \pi_- \right], \quad 
 \hat Y  =   -\frac{i}{2}\left[ X_{+} -    X_{-} \right] 
 - \frac{1}{2M\Omega}\left[ \pi_{+} +    \pi_- \right], 
\eeq

\beq
\hat P_{x}  = -\frac{i}{4}M\Omega\left[ X_{+} -    X_{-}\right] 
 + \frac{1}{4}\left[\pi_{+}  +    \pi_- \right], \quad 
 \hat P_{y}  =   -\frac{1}{4}M\Omega\left[X_{+} +   X_{-}\right] 
 -  \frac{i}{4}\left[\pi_{+}  -   \pi_- \right].
\eeq

%
%

The mean values of $\hat X, \hat Y$ and $\hat P_{x}, \hat P_{y}$ in the states  $|z\rangle_{m}$ are given in terms of the 
classical dynamical variables $p, q$ with $z = \frac{q + ip}{\sqrt{2}}$ and the modified Bessel functions of the first kind,  
where   the notation 
$\xi_{a, b}(z) = I_a(2|z|)/I_b(2|z|)$ has been introduced,  
 by 
\beq
   \langle \hat X \rangle_{z,m}  =  
   \sqrt{\frac{\hbar}{ M\Omega}}q \, \sqrt{\frac{\xi_{m, m-1}(z)}{|z|}},
 \eeq 


  \beq
    \langle \hat Y\rangle_{z,m}   =    \sqrt{\frac{\hbar}{2M\Omega}}\left[p\, \sqrt{\frac{2\xi_{m, m-1}(z) }{|z|}} 
- 2 \sqrt{|z|\xi_{m, m+1}(z)} \right],
  \eeq
 
  \beq
   \langle \hat P_{x} \rangle_{z,m}    =   
\frac{1}{2}\sqrt{\frac{\hbar M \Omega}{2}}\left[p\, \sqrt{\frac{2\xi_{m, m-1}(z)}{|z|}} 
 + 2 \sqrt{|z|\xi_{m, m+1}(z)} \right],
  \eeq
  

  \beq
  \langle \hat P_{y} \rangle_{z,m}    =  
  -\frac{1}{2}\sqrt{\frac{\hbar M \Omega}{2}} q\,
    \sqrt{\frac{2\xi_{m, m-1}(z)}{|z|}}.
\eeq

Thereby, 

\beq
 \langle \hat P_{y} \rangle_{z,m}  = -\lambda  \langle \hat X \rangle_{z,m}, \,\; 
    {\langle \hat P_{x} \rangle_{z,m}} = 
   \lambda   \langle \hat Y\rangle_{z,m}  + \sqrt{2\hbar M \Omega}\sqrt{|z|\xi_{m, m+1}(z)}, \quad  \lambda = \frac{M \Omega}{2}. 
\eeq



%
%
%
%

  Moreover, we obtain for $\hat X \hat P_{x} $ and $\hat Y \hat P_{y}:$ 
  
  \beq
  \langle \hat X \hat P_{x} \rangle_{z,m}  &=&  
  \frac{\hbar}{2} \, \frac{pq}{|z|} \sqrt{\xi_{m, m-2}(z)} +  \frac{i \hbar}{2} +  
   \frac{\hbar \sqrt{2}}{2}q,\crcr
 &&\crcr &&\crcr 
  \langle \hat Y \hat P_{y} \rangle_{z,m}  & = &  
  -\frac{\hbar}{2} \, \frac{pq}{|z|}\sqrt{\xi_{m, m-2}(z)}
   +  \frac{i \hbar}{2} +  
   \frac{\hbar \sqrt{2}}{2}q.
  \eeq
  
  The observable dispersions are derived with the approximations as performed in (\ref{approxim})-(\ref{approx}), 
  and lead,  for $|z|\gg 1, $ to the following relation:
%
%
%
%
%
 \beq\label{uncert00}
(\Delta \hat X)^2_{z,m}  (\Delta \hat P_x)^2_{z,m}   
     &=& \frac{\hbar^2}{4}(m+1)^2  + \left(\frac{4\hbar }{M\Omega}+ \frac{2\hbar}{M\Omega}\frac{1}{|z|}- \frac{\hbar^2}{2} \right)\crcr
     && \times \left(\sqrt{2}(m+1)p  +  \frac{2}{|z|}p^2\right)
    +   \frac{\hbar^2}{4}\frac{\sqrt{2}(m+ 1)}{|z|}p.
 \eeq
 
 The relation (\ref{uncert00}) yields,  in the limit $p \rightarrow 0, $ 
 \beq
 (\Delta \hat X)^2_{z,m}  (\Delta \hat P_x)^2_{z,m}  = \frac{\hbar^2}{4}(m+1)^2 \geq \frac{\hbar^2}{4}, \qquad  m \geq 0,  
 \eeq
 satisfying  
 $(\Delta \hat X)_{z,m}  (\Delta \hat P_x)_{z,m}  = \frac{\hbar}{2} $ at $m=0, $ which means that the uncertainty relation is saturated, 
 highliting one of the most important features exhibited by   
 CS, i.e, their quantal face is the closest possible to its classical counterpart.

  

\subsection{Probability density and time evolution}

This paragraph explores   the semi-classical character of the   $\mathfrak{su}(1,1)$ CS $|z\rangle_{m}. $ We analyse  
how these states  do evolve in time under the action of the time evolution operator provided by 
the physical Hamiltonian describing the quantum system. 


From the quantity 

\beq
_{m}\langle z'|z\rangle_{m} 
&=& \left(\frac{|z'z|}{{\bar{z'}}z}\right)^{m/2} \frac{1}{\sqrt{I_{m}(2|z'|)I_{m}(2|z|)}}
\sum_{\upsilon=0}^{+\infty}\frac{(\sqrt{ \bar z' z})^{2\upsilon +m}}{\Gamma(\upsilon+1)\Gamma(\upsilon+m+1)},   
\eeq

given a  normalized state $|z_0\rangle_{m}$,    the phase space distribution is defined by   the probability density  
 as follows:

\beq\label{probadens}
z \mapsto \varrho_{z_0}(z) :=   |_{m}\langle z|z_0\rangle_{m}|^2
&=& \frac{I_m(2\sqrt{z_0 \bar z}) I_m(2\sqrt{\bar z_0  z})}{I_{m}(2|z|)I_{m}(2|z_0 |)}.
\eeq

Its time evolution behavior
is then provided by 

\beq
z \mapsto \varrho_{z_0}(z,t) := |_{m}\langle z|e^{-\frac{i}{\hbar} \tilde H t}|z_0\rangle_{m}|^2.
\eeq

By acting the evolution operator $U(t) = e^{-\frac{i}{\hbar} \tilde H t}$ on the state $|z_0\rangle_{m} $,  we get
\beq 
| z_0; t\rangle_{m}  &=&  e^{-\frac{i}{\hbar} \tilde H t}|z_0\rangle_{m}  \cr
&=&  e^{-\frac{i}{2}[m(\Omega + \omega_{c}) + \Omega]t}\left[\frac{|z_0|^{m/2}}{\sqrt{I_{m}(2|z_0|)}}\sum_{\nu=0}^{+\infty}\,
\frac{(z_0 (t))^{\nu}}{\sqrt{\Gamma(\nu+1)\Gamma(\nu+m+1)}}|\nu,m\rangle \right] \cr
&=& 
e^{-\frac{i}{2}[m(\Omega + \omega_{c}) + \Omega]t}| z_0(t)\rangle_{m}
\eeq
where $z_0 (t) :=  e^{-i\Omega t} z_0.$


Therefore

\beq
\varrho_{z_0}(z,t) := 
|_{m}\langle z|e^{-\frac{i}{\hbar} \tilde H t}|z_0\rangle_{m}|^2  = 
\frac{I_m(2\sqrt{z_0 (t)\bar z}) I_m(2\sqrt{\bar z_0(t) z})}{I_{m}(2|z|)I_{m}(2|z_0(t)|)}.
\eeq

It comes that the time dependence of a given BGCS $|z\rangle_{m}$ is furnished by 

\beq{\label{timeev}}
| z; t\rangle_{m} = e^{-\frac{i}{\hbar} \tilde H t}|z\rangle_{m}  = e^{-\frac{i}{2}[m(\Omega + \omega_{c}) + \Omega]t}| z(t)\rangle_{m}, 
\quad z (t) := e^{-i\Omega t} z.
\eeq

The relation (\ref{timeev}) shows that the time evolution   of the CS $|z\rangle_{m}$ reduces to  a rotation in the complex plane 
given by $z\mapsto  z (t) = e^{-i\Omega t} z$ up to a phase, namely, 
$e^{-\frac{i}{2}[m(\Omega + \omega_{c}) + \Omega]t}$.
 Therefore, the semi-classical feature of the CS is given by (\ref{probadens}), while 
 the temporal stability property  is highlighted by the relation (\ref{timeev}).  The latter asserts
that the temporal evolution of any CS always remains a CS, 
and fixes the phase
behavior of the CS $|z\rangle_{m}$ with the factor $e^{-i\Omega t} .$


\section{Quantization with the $\mathfrak{su}(1,1)$ coherent states}

As is
proved in  Section $3$, the  $\mathfrak{su}(1,1)$ CS family  resolves the unity. As an immediate consequence, 
we establish in this section the correspondence (quantization) between classical and quantum quantities. For more details in the quantization 
procedure see  \cite{gazbook09, agh} and references listed therein.

\subsection{Quantization of elementary classical observables}
The Berezin-Klauder-Toeplitz quantization of elementary classical variables $z$ and $\bar z$ is realized via the maps
$z \longmapsto A_z$ and $\bar z \longmapsto  A_{\bar z}$  defined on the Hilbert subspaces $\mathfrak H_{m}$ and 
$\mathfrak H'_{m'}$ by

\beq
A_{z_{|\mathfrak H_{m}}}   := \int_{\mathbb C}z \, |z\rangle_{m}\, _{m}\langle z|d\varrho(z), \qquad 
\quad A_{z_{|\mathfrak H'_{m'}}}   := \int_{\mathbb C}z \, 
|z\rangle_{m'}\, _{m'}{\langle z}|d\tilde\varrho(z)
\eeq 

providing


\beq
A_{z}   := \sum_{m' < 0}
\int_{\mathbb C}z \, 
|z\rangle_{m'}\, _{m'}{\langle z}|d\tilde\varrho(z)  \bigoplus
\sum_{m=0}^{+\infty} \int_{\mathbb C}z \, |z\rangle_{m}\, _{m}\langle z|d\varrho(z), 
\eeq

\beq
A_{\bar z}   := \sum_{m' < 0}
\int_{\mathbb C}\bar z \, 
|z\rangle_{m'}\, _{m'}{\langle z}|d\tilde\varrho(z)  \bigoplus \sum_{m=0}^{+\infty}
\int_{\mathbb C}\bar z \, |z\rangle_{m}\, _{m}\langle z|d\varrho(z).
\eeq

This gives,
 using the equations  (\ref{rep2}),   (\ref{gen11}) and  (\ref{gen12}), the following  relations:

\beq\label{quant02}
A_z  
&=& \sum_{m' < 0}
\sum_{n=0}^{+\infty}\sqrt{(n-m'+1)(n+1)} |n, m'\rangle \langle n+1,m'|
\crcr
 && \bigoplus \sum_{m=0}^{+\infty}
\sum_{n=m}^{+\infty}\sqrt{(n-m+1)(n+1)} |n,m\rangle \langle n+1,m|= \mathcal K_{-},
\eeq

\beq\label{quant03}
A_{\bar z}  
&=& \sum_{m' < 0}
\sum_{n=0}^{+\infty}\sqrt{n(n-m')} |n, m'\rangle \langle n-1, m'|\crcr
 && \bigoplus \sum_{m=0}^{+\infty}
\sum_{n=m}^{+\infty}\sqrt{n(n-m)} |n,m\rangle \langle n-1,m|= \mathcal K_{+}
\eeq


with the matrix elements:
\beq
(A_{z_{|\mathfrak H_{m}}})_{n,k} &=& \frac{2}{\pi} \int_{0}^{\infty} K_{m}(2\rho) 
\rho^{n-m+k+2}  d\rho   \times \int_{0}^{2\pi}e^{-\imath[(k-m)-(n+1-m)]\phi}d\phi \times \cr
&& \times \frac{1}{\sqrt{\Gamma(n-m+1)\Gamma(n+1)\Gamma(k-m+1)\Gamma(k+1)}},   
\eeq

\beq
(A_{\bar z_{|\mathfrak H_{m}}})_{n,k} &=& \frac{2}{\pi} \int_{0}^{\infty} K_{m}(2\rho) 
\rho^{n-m+k+2}  d\rho   \times \int_{0}^{2\pi}e^{-\imath[(k+1-m)-(n-m)]\phi}d\phi \times \cr
&& \times \frac{1}{\sqrt{\Gamma(n-m+1)\Gamma(n+1)\Gamma(k-m+1)\Gamma(k+1)}},   
\eeq
respectively.

%
%


Their commutator $[A_z,  A_{\bar z}] $ is reduced to
\beq\label{opcom00}
[A_z,  A_{\bar z}]  =   2\mathcal K_3  
\eeq
 reminding the $\mathfrak{su}(1,1)$ commutation rules (\ref{eq23}).

%
%
%
%
%
%
%
%
%
%
%
%
%
%
%

Other interesting  results emerging from this context are the following  mean values: 

\beq\label{meanv00}
_{m}\langle z| A_{z_{|\mathfrak H_{m}}}|z\rangle_{m} = z = \, _{m'}{\langle z}|
 A_{z_{|\mathfrak H'_{m'}}}|z\rangle_{m'},
\,\, _{m}\langle z|  A_{\bar z_{|\mathfrak H_{m}}} |z\rangle_{m} = \bar z 
= \, _{m'}{\langle z}| A_{\bar z_{|\mathfrak H'_{m'}}}|z\rangle_{m'},
\eeq

\beq
_{m}\langle z|  A^{2}_{z_{|\mathfrak H_{m}}}|z\rangle_{m} =  z^2 
= \, _{m'}{\langle z}| A^{2}_{z_{|\mathfrak H'_{m'}}}|z\rangle_{m'}, \,\, 
_{m}\langle z|  A^{2}_{\bar z_{|\mathfrak H_{m}}} |z\rangle_{m} = {\bar z}^2 = \, _{m'}{\langle z}|A^{2}_{\bar z_{|\mathfrak H'_{m'}}}|z\rangle_{m'}, \nonumber \\
\eeq
\beq\label{meanv01}
 _{m}\langle z|  A_{\bar z_{|\mathfrak H_{m}}}A_{z_{|\mathfrak H_{m}}}|z\rangle_{m} 
 = |z|^2 =  \, _{m'}{\langle z}|A_{\bar z_{|\mathfrak H'_{m'}}}A_{z_{|\mathfrak H'_{m'}}} 
 |z\rangle_{m'}
 \eeq
 
 \beq
 _{m}\langle z|  A_{z_{|\mathfrak H_{m}}} A_{\bar z_{|\mathfrak H_{m}}} |z\rangle_{m} = |z|^2 + 
2 \langle \mathcal K_{3}\rangle_{z,m}, \,\, 
_{m'}{\langle z}| A_{z_{|\mathfrak H'_{m'}}}\,A_{\bar z_{|\mathfrak H'_{m'}}}|z\rangle_{m'} 
= |z|^2 + 2\langle \mathcal K_{3}\rangle_{z,m'} \nonumber \\
\eeq
where $\langle \mathcal K_{3}\rangle_{z,m}$ is provided in (\ref{gen14}) and $\langle  \mathcal K_{3}\rangle_{z,m'} $ given by 
 
 \beq 
\langle \mathcal K_{3}\rangle_{z,m'}  =  
|z| \frac{I_{-m'+1}(2|z|)}{I_{-m'}(2|z|)} + \frac{-m'+1}{2}.
\eeq

\section{Mandel parameter}

Several parameters can be introduced to characterize  statistical properties. The most popular one is  the Mandel parameter 
 \cite{mandel1, mandel-wolf}. Let   $\mathcal Q$ denote this parameter. Then $\mathcal Q$ is
a convenient noise-indicator of a non-classical field. This is
  frequently used to measure the deviation from Poisson distribution. Thus, $\mathcal Q$ can distinguish quantum processes from 
classical ones \cite{mandel1}-\cite{lietal}.   

The Mandel parameter  $\mathcal Q$ is defined as \cite{mandel1} 

\beq
\mathcal Q &\equiv& \frac{ (\Delta N)^{2} -  \langle N \rangle}{\langle N \rangle} = \frac{2\langle I \rangle}{T}\int_{0}^{T} dt_2 \int_{0}^{t_2}dt_1[1 + \lambda(t_1)]  - \langle I \rangle T,
\eeq
where  $\langle N \rangle$ is the average counting number;
 $ (\Delta N)^{2}$ is the corresponding  square  variance;
 $\langle I \rangle = \langle N \rangle / T$ is the steady-state photon-counting rate expressed in units of cps;
$\lambda(\tau) = \langle \Delta I(t)\Delta I(t+\tau) \rangle /\langle I(t) \rangle \langle I(t+\tau) \rangle$ is the normalized 
two-time correlation 
of intensity  fluctuations $(\Delta I(t) = I(t) - \langle I(t) \rangle)$ of time difference equal to the time  $\tau$ \cite{mandel2}.

Moreover, the Mandel parameter $\mathcal Q$

\beq
\mathcal Q = \frac{(\Delta N)^{2} }{\langle N \rangle} - 1 \equiv \mathcal F - 1
\eeq
 is closely related to the normalized variance,  also called the quantum Fano factor $\mathcal F$ \cite{bajer-miranowicz}, 
  given   by $\mathcal F =  (\Delta N)^{2} /\langle N \rangle$, of the photon distribution. For 
$\mathcal F < 1 (\mathcal Q \leq 0)$, the emitted light 
  is referred to as  sub-Poissonian (corresponding to nonclassical states);   $\mathcal F = 1, \mathcal Q = 0$ corresponds to 
 the Poisson distribution (case of standard CS), 
whereas for $\mathcal F > 1, (\mathcal Q > 0)$ the 
  light is called super-Poissonian (corresponding to  classical states) \cite{mandel1}-\cite{zhangetal}.  Thus, in this case, using 
 (\ref{numb00}), we have in the BGCS $|z\rangle_{m}$:

\bitem

\item For $|z| \ll 1$

\beq
(\Delta N)^2_{z,m}  &\simeq& 
|z|^{2} \frac{|z|^{2}}{(m+2)(m+1)}+  |z|\frac{|z| }{m+1}   - |z|^{2}\frac{|z|^{2}}{(m+1)^{2}}, \cr
\langle N \rangle_{z,m} &\simeq& |z|\frac{|z|}{m+1} 
\eeq

providing $\mathcal F <1$ and

\beq
\mathcal Q   \simeq  -\frac{|z|^{2}}{(m+1)(m+2)}<0
\eeq
indicating that the BGCS $|z\rangle_{m}$ have   sub-Poissonian statistics 
 as discussed in the subsection \ref{paragraph}

\item  For $|z| \gg 1$

\beq
 (\Delta N)^2_{z,m}  \simeq |z|, \qquad \langle N \rangle_{z,m}   \simeq |z|
\eeq

giving $\mathcal F \simeq 1$ such that 

\beq
\mathcal Q   \simeq 0.
\eeq
It implies  that  the BGCS $|z\rangle_{m}$ have Poissonian statistics for $|z| \gg 1$ as also noticed in the subsection \ref{paragraph} 
 Therefore,  for $|z| \gg 1$, the sates  $|z\rangle_{m}$ coincide with the standard CS. 
\eitem

\section{Thermal properties of the  BGCS}

In quantum mechanics, the   probability distribution  on the states of  a physical system  can be characterized by 
a statistical operator called  density matrix,  generally denoted by $\rho$. The latter reveals to be an important tool 
used for 
examining the  physical and chemical properties of  a system. In the Fock representation, 
the density matrix is given by its matrix 
elements, $\rho_{m,n} = \langle m|\rho|n\rangle$, as  in
\cite{gazbook09}: 
$\rho =  \sum_{m,n}  \rho_{m,n} |m\rangle \langle n|$. 

This section furnishes a description of  the statistical properties of the BGCS for our model in the situation  of a pseudo-thermal
equilibrium (limited to some value
of $m$, with $m \geq 0$).
Consider a quantum gas of 
the system in the 
thermodynamic equilibrium with a reservoir  at temperature $T$, which satisfies a quantum canonical distribution. 
The corresponding normalized density operator for a fixed number $m$ is given by

\beq{\label{stat00}}
\rho_{m}= \frac{1}{Z}\sum_{\nu=0}^{+\infty}e^{-\beta E_{\nu, m}}|\nu,m\rangle \langle \nu,m|, 
\quad E_{\nu, m} = \hbar \Omega \nu +\frac{\hbar}{2}\left(m(\Omega + \omega_{c}) + \Omega\right),
\eeq
where the partition function $Z$ is taken as the normalization constant.
To derive the energy levels  $E_{\nu, m}$ in (\ref{stat00}), 
 we put $n= \nu + m$ in $\mathcal E_{n, m}$ given in (\ref{equ11})  and find

\beq
\mathcal E_{n,m}  =  
 \hbar \Omega \left(\nu + m + \frac{1}{2}\right)  - \frac{\hbar}{2} \Omega m + 
\frac{\hbar}{2} \omega_c m. 
\eeq

Developing and rearranging this expression lead to the required relation, i.e.,

\beq
\mathcal E_{n,m} =  \hbar \Omega \nu + \frac{\hbar}{2} \left[\Omega + (\Omega  + \omega_c) m \right]:= 
E_{\nu, m}.
\eeq


The diagonal elements of the density operator in  the BGCS representation are given by
\beq
_{m}\langle z|\rho_{m}|z \rangle_{m} = \frac{1}{Z}\frac{|z|^{m}}{I_{m}(2|z|)}\sum_{\nu=0}^{+\infty}\, e^{-\beta E_{\nu, m}}
\frac{|z|^{2\nu}}{\Gamma(\nu+1)\Gamma(\nu+m+1)}.
\eeq
Note that the quantity $_{m}\langle z|\rho_{m}|z \rangle_{m}$ is analog to the \textquotedblleft semi-classical \textquotedblright phase space 
distribution 
function  $\mu(x,p) =  \langle z|\rho|z \rangle$ 
 associated to the density matrix $\rho$ (here $\rho_{m}$)  of the system which is normalized as $\int(dx dp/2\pi \hbar)\mu(x,p) = 1.$ It 
is often referred to 
as the Husimi distribution \cite{husimi}.   

Then, using (\ref{rec01}), (\ref{stat00}) and 
$E_{\nu, m} = 
\hbar \Omega \nu +\frac{\hbar}{2}\left(m(\Omega + \omega_{c}) + \Omega\right)$, we deduce
\beq{\label{stat000}}
_{m}\langle z|\rho_{m}|z \rangle_{m} = \frac{1}{Z}e^{-\frac{\beta \hbar}{2}\Omega} 
e^{-\frac{\beta \hbar}{2}m \omega_{c} }\,\frac{I_{ m}(2|z|
e^{-\frac{\beta \hbar}{2} \Omega })}{I_{m}(2|z|)}.
\eeq
The normalization of the density operator leads to
\beq
 Tr \rho_{m} = \int_{\mathbb C} d\varrho(z)\, _{m}\langle z|\rho|z \rangle_{m} = 1.
\eeq
By the use of the following integral relation \cite{gradshtein}:
\beq{\label{int_00}}
&&\int_{0}^{+\infty} dx \, x^{-\lambda}K_{\mu}(ax)I_{\upsilon}(bx) 
= \frac{b^{\upsilon}\Gamma\left(\frac{1}{2} - \frac{1}{2}\lambda + \frac{1}{2}\mu + 
\frac{1}{2}\upsilon\right)\Gamma\left(\frac{1}{2} - \frac{1}{2}\lambda - \frac{1}{2}\mu + 
\frac{1}{2}\upsilon\right)}{2^{\lambda + 1}\Gamma(\upsilon+1)a^{-\lambda + \upsilon + 1}}\crcr
&&\times F\left(\frac{1}{2} - \frac{1}{2}\lambda + \frac{1}{2}\mu + \frac{1}{2}\upsilon, \frac{1}{2} - 
\frac{1}{2}\lambda - \frac{1}{2}\mu + \frac{1}{2}\upsilon;\upsilon + 1;\frac{b^{2}}{a^{2}}\right)\crcr
&&[Re (\upsilon + 1 - \lambda \pm \mu) > 0, a > b],
\eeq
we obtain 
\beq{\label{numb02}}
Z= e^{-\frac{\beta \hbar}{2}[m(\Omega + \omega_{c}) + \Omega]}F\left(m+1,1;m+1;
e^{-\beta \hbar\Omega }\right),
\eeq
where $F$ corresponds to the hypergeometric function  satisfying the following property \cite{nikiforov}:
\beq{\label{numb03}}
F(c,\mu;c;x) = F(\mu,c;c;x) = (1-x)^{-\mu}.
\eeq
It comes that the partition function $Z$ takes the  form:
\beq{\label{stat03}}
Z =  e^{-\frac{\beta \hbar}{2}m(\Omega + \omega_{c})}
\left\{2\sinh\left\{\frac{\beta \hbar}{2}\Omega  \right\} \right \}^{-1}.
\eeq
The diagonal elements of the density matrix (\ref{stat000}) can be written as
\beq\label{density00}
_{m}\langle z|\rho_{m}|z \rangle_{m} = 2e^{\frac{\beta \hbar}{2}   (m-1)\Omega}
\sinh\left\{\frac{\beta \hbar}{2} \Omega  \right\} \,\frac{I_{m}(2|z|
e^{-\frac{\beta \hbar}{2} \Omega })}{I_{m}(2|z|)}.
\eeq
In  the case of a strong magnetic field, i.e. $\omega_{0} \ll \omega_{c}$, it simplifies into the expression:
\beq\label{exppart}
_{m}\langle z|\rho_{m}|z \rangle_{m}  &=& 
2e^{\frac{\beta \hbar \omega_{c}}{2}(m-1)\left(1+2\left(\frac{\omega_{0}}{\omega_{c}}\right)^{2}\right)}
\, \sinh\left\{\frac{\beta \hbar \omega_{c}}{2}\left(1+2\left(\frac{\omega_{0}}{\omega_{c}}\right)^{2}\right)  \right\}\crcr 
&&\times 
\frac{I_{m}\left(2|z|
e^{-\frac{\beta \hbar \omega_{c}}{2}\left(1+2\left(\frac{\omega_{0}}{\omega_{c}}\right)^{2}\right)}\right)}{I_{m}(2|z|)}.
\eeq

With the equations (\ref{density00}) and (\ref{exppart}), we look for
  another 
thermodynamical aspect of the  model in the BGCS representation. Indeed,   
CS are also relevant to the concept of Wehrl entropy $W$ defined as \cite{wehrl, anderson-halliiwell,lieb}:

\beq
W: = -\int \frac{dx dp}{2\pi \hbar} \mu(x,p) \ln \mu(x,p)
\eeq

where $\mu(x,p) =  \langle z|\rho|z \rangle$.  Evaluating it now in terms of the  distribution 
function $_{m}\langle z|\rho_{m}|z \rangle_{m}$   
in (\ref{density00}), we obtain

\beq\label{ent00}
W &=& -\int_{\C} d\varrho(z)\, _{m}\langle z|\rho_{m}|z \rangle_{m} 
\, \ln[_{m}\langle z|\rho_{m}|z \rangle_{m}] \cr
&=&  - \ln \left\{2
\sinh\left\{\frac{\beta \hbar}{2} \Omega \right\} \right\}  -\frac{\beta \hbar}{2} \Omega (m-1) 
-\left[2e^{\frac{\beta \hbar}{2} \Omega (m-1)}
\sinh\left\{\frac{\beta \hbar}{2} \Omega  \right\} \right] \times \cr
&& \times \int_{\C} d^{2}z \frac{2}{\pi}K_{m}(2|z|)\,I_{m}(2|z|e^{-\frac{\beta \hbar}{2} \Omega })
\,\ln \left\{
\frac{I_{m}(2|z|e^{-\frac{\beta \hbar}{2} \Omega })}{I_{m}(2|z|)}\right\}. 
\eeq

Assuming that the argument of the logarithm function under the integral  is dominated in the region $|z| \ll 1$, we get approximately

\beq
\int_{\C} d^{2}z \frac{2}{\pi}K_{m}(2|z|)\,I_{m}(2|z|e^{-\frac{\beta \hbar}{2} \Omega })
\,\ln \left\{
\frac{I_{m}(2|z|e^{-\frac{\beta \hbar}{2} \Omega })}{I_{m}(2|z|)}\right\} 
\simeq  -m\frac{\beta \hbar}{2} \Omega  
\times \frac{e^{-\frac{\beta \hbar}{2} \Omega (m-1)}}{2\sinh\left\{\frac{\beta \hbar}{2} \Omega  \right\}}.
\eeq

Therefore,  the  Wehrl entropy is approximated to the quantity

\beq\label{ent01}
W 
&\simeq&  -\ln \left[1-e^{-\beta \hbar \Omega }\right] = -1 + W_{HO},
\eeq 
where $ W_{HO} = 1 - \ln \left[1-e^{-\beta \hbar \Omega  }\right]$ is 
the conventional harmonic oscillator   Wehrl entropy \cite{anderson-halliiwell} 
with   frequency $\Omega$. Besides,  setting 

\beq\label{denst}
\tilde{_{m}\langle  z| \rho|z \rangle_{m}}  = \left(\frac{2\pi l^{2}}{A}\right)\, _{m}\langle z|\rho|z \rangle_{m}
\eeq

where $l = \sqrt{\frac{\hbar}{M \Omega  }}$ 
and $A = \pi \mathcal R^2$ with $\mathcal R$ the radius of the cylindrical body 
considered in \cite{feldman-kahn},  
and  evaluating

\beq
\tilde W = -\frac{A}{2\pi l^{2}}\int_{\mathbb C} d\varrho(z)\, \tilde{_{m}\langle  z| \rho|z \rangle_{m}}
\ln [\tilde{_{m}\langle  z| \rho|z \rangle_{m}}]
\eeq

 as in (\ref{ent00})-(\ref{ent01}), we get 

\beq
\tilde W \simeq -\ln \left[1-e^{-\beta \hbar \Omega  }\right] - 
\ln \left(\frac{2\pi l^{2}}{A}\right) \equiv -1 + W_{calc}. 
\eeq 
Here, $ W_{calc} = 1-\ln \left[1-e^{-\beta \hbar \Omega}\right] - 
\ln \left(\frac{2\pi l^{2}}{A}\right)$ is the Wehrl entropy calculated for Landau's diamagnetism for a spinless electron in a 
uniform magnetic  field.   
 The thermal harmonic oscillator  \cite{pennini et al} frequency is  denoted  by $\Omega$,  and $l = \sqrt{\frac{\hbar}{M \Omega}}$. 
Hence, in this case, our physical model  is an approximation 
of  the problem of a thermal harmonic oscillator with frequency $\Omega$.

In the case of a strong magnetic field, using (\ref{exppart}),  we get

\beq
W \simeq -\ln \left(1-e^{-\frac{\beta \hbar \omega_{c}}{2}\left(1+2\left(\frac{\omega_{0}}{\omega_{c}}\right)^{2}\right)}\right).
\eeq
 
The diagonal expansion of the density operator can be performed for this physical model, 
as investigated in \cite{brif-aryeh} in the constructed BGCS as follows:
\beq
\rho_{m} = \int_{\mathbb C} d\varrho(z) P_{m}(z)|\nu, m\rangle \langle \nu,m|,
\eeq
where the function $P_{m}(z)$ must be determined. To this end, we first compute 
the diagonal elements of the density operator $\rho_{m}$ in the basis of the number states, namely  
$\{|\nu, m\rangle, \nu \geq 0\}$, by setting:
\beq{\label{stat06}}
\langle \nu,m|\rho_{m}|\nu, m\rangle = \int_{\mathbb C} 2\frac{d^{2}z}{\pi}K_{m}(2|z|)I_{m}(2|z|)
P_{m}(z)\langle \nu,m|z \rangle_{m}\, _{m}\langle z|\nu, m\rangle.
\eeq
By the use of $\sum_{\nu = 0}^{+\infty}|\nu, m\rangle \langle \nu,m| = 1$ 
and (\ref{stat00}), they are given by
\beq
\langle \nu,m|\rho_{m}|\nu, m\rangle = \frac{1}{Z}e^{-\beta E_{\nu, m}} = [1 - 
e^{-\beta \hbar\Omega }]e^{- \beta \hbar \Omega  \nu}.
\eeq

Then, using the integral (\ref{stat007}), we obtain for the function $P_{m}(z)$ the following 
expression:
\beq{\label{stat09}}
P_{m}(z) = [e^{\beta \hbar \Omega } - 1]e^{\frac{\beta \hbar}{2}\Omega m}
\frac{K_{m}(2|z|e^{\frac{\beta \hbar}{2} \Omega })}{K_{m}(2|z|)}
\eeq

which is normalized as 
\beq
\int_{\mathbb C} d\varrho(z)P_{m}(z) = 1.
\eeq

Then, the diagonal representation of the normalized density operator takes the form
\beq
\rho_{m} = [e^{\beta \hbar \Omega  } - 1]e^{\frac{\beta \hbar}{2} \Omega m}
\int_{\C} d\varrho(z)\frac{K_{m}(2|z|e^{\frac{\beta \hbar}{2}\Omega })}{K_{m}(2|z|)} 
|z\rangle_{m} \, _{m}\langle z|.
\eeq
 
Therefore, given an observable $\mathcal O$, one obtains its mean value, i.e.,  its pseudo-thermal 
average (with $m \geq 0$)  as follows:
\beq{\label{num05}}
\langle \mathcal O \rangle_{m} = Tr (\rho_{m} \mathcal O) = \int_{\mathbb C} d\varrho(z) P_{m}(z) \,_{m}\langle z|\mathcal O |z\rangle_{m}.
\eeq

 In this manner,  the {\it pseudo}-thermal
 expectation value of the number operator $N$, by using (\ref{numb00}), (\ref{numb02}) and (\ref{numb03}) 
 together, is
\beq
\langle N \rangle_{m} = Tr (\rho_{m} N) = \int_{\C} d\varrho(z) P_{m}(z) \,_{m}\langle z|N|z\rangle_{m} = 
\frac{1}{e^{\beta \hbar \Omega} - 1}.
\eeq

%

In the same vein, using (\ref{numb00}), (\ref{numb02}) and (\ref{numb03}),  
 the {\it pseudo}-thermal  
expectation value of the square of the number operator becomes
\beq
\langle N^{2} \rangle_{m} = Tr (\rho_{m} N^{2}) = \int_{\mathbb C} d\varrho(z) P_{m}(z) \,_{m}\langle z|N^{2}|z\rangle_{m} = \frac{1}{e^{\beta \hbar \Omega} - 1} + 2 \frac{1}{(e^{\beta \hbar \Omega} - 1)^{2}}.
\eeq
One remarks that both  thermal expectation values $\langle N \rangle_{m}$ and  $\langle N^{2} \rangle_{m}$ 
are independent of the Bargmann index given here by $m$.
One can therefore define the thermal intensity correlation function, which is also independent of the index $m$, by
\beq
\langle g \rangle_{m} \equiv \frac{\langle N^{2} \rangle - \langle N \rangle}{\langle N \rangle^{2}} = \langle g^{2} \rangle = 2.
\eeq

%


\section{Concluding remarks}

In this work, we   investigated the Fock-Darwin Hamiltonian describing a gas of spinless charged particles, subject to a 
perpendicular magnetic field ${\bf B},$  confined in a harmonic potential.
We used a set of  step and orbit-center coordinate operators. Then we   showed that the studied system possesses   
$\mathfrak{su}(1,1)$ Lie algebra. 
 As a consequence, CS were constructed   as the 
eigenstates of the {\rm SU}$(1,1)$ group generator $\mathcal K_{-}.$
 The mean
values of {\rm SU}$(1,1)$ group generators and of the physical system observables, the probability density and the time dependence of the
BGCS were discussed. Using 
these CS,   the Berezin - Klauder - Toeplitz quantization
  were 
 performed. 
Quantum  optical characteristics   such as   the 
Mandel parameter were derived.
 Statistical properties of a gas in  
thermodynamic equilibrium with a reservoir  at temperature $T$,  satisfying the quantum canonical distribution, 
were investigated and discussed. Finally, using the density matrix provided in the BGCS  representation,
 we  noticed, via the calculation of the Wehrl entropy,  that the considered system can be identified to a model of
a thermal harmonic oscillator.

\section*{Acknowledgments}
The authors are grateful to anonymous referees for their useful comments which permit to 
substantially improve the paper. MNH thanks Professor G. A. Goldin from Rutgers University (USA) for fruitful discussions and suggestions. This work is partially supported by the ICTP through the  OEA-ICMPA-Prj-15. The ICMPA is in partnership with the Daniel
Iagolnitzer Foundation (DIF), France.


\begin{thebibliography}{10}
\addcontentsline{toc}{chapter}{References}
\bibitem{manko}
I. A. Malkin and V. I. Man'ko: \emph{Coherent States of a Charged Particle in a
Magnetic Field}, {\em Zh. Eksp. Teor. Fiz.\/} {\bf 55}, 1014 (1968).
\bibitem{feldman-kahn}
A. Feldman and A. H. Kahn: 
\emph{Landau diamagnetism from the coherent states of an electron in a uniform magnetic field},
 {\em Phys. Rev. B\/}  {\bf 1}, 4584 (1970).
\bibitem{gazeau-hsiao-jellal}
J. P. Gazeau,  P. Y. Hsiao and A. Jellal: 
\emph{A Coherent-State Approach to Two-dimensional Electron Magnetism}, 
 {\em Phys. Rev. B\/}  {\bf 65}, 094427 (2002).
\bibitem{schuch-moshinsky}
D. Schuch and M. Moshinsky: \emph{Coherent states and dissipation for the motion of a charged particle in a constant 
magnetic field}, 
{\em J. Phys. A: Math. Gen.\/}  {\bf 36}, 6571 (2003).
\bibitem{gazeau-novaes}
J. P. Gazeau and M. Novaes:
\emph {Multidimensional generalized coherent states,} 
{\em J. Phys. A: Math. Gen. 
\/}  {\bf  36}, 199-212
(2003).
\bibitem{perelomov}
A. M. Perelomov: \emph{Generalized Coherent States and Their Applications\/},  Springer, Berlin 1986. 
\bibitem{landau2}
L. D. Landau: \emph{Diamagnetismus der Metalle, } {\em  Z. Phys.\/}   {\bf 64}, 629 (1930).
\bibitem{fakhri1}
H. Fakhri:
\emph{Generalized Klauder-Perelomov and Gazeau-Klauder coherent states
for Landau levels}, 
 {\em Phys. Lett. A\/}  {\bf 313},  243-251 (2003).
\bibitem{antoine-gazeau-klauder}
J. P. Antoine, J. P. Gazeau, P. Monceau, J. R. Klauder and K. A. Penson:
\emph{Temporally stable coherent states for infinite well and P\"{o}schl-Teller potentials},
 {\em J. Math. Phys.\/} {\bf  42}, 2349-2387 (2001).
\bibitem{fakhri3}
H. Fakhri: 
\emph{$su(1,1)$-Barut-Girardello coherent states for Landau levels},  
{\em J. Phys. A: Math. Gen.\/} {\bf 37},  5203-5210 (2004).
\bibitem{barut-girardello}
A. O. Barut and L. Girardello:  \emph{New
\textquotedblleft coherent\textquotedblright  states associated with non
compact groups, }   {\em Commun. Math. Phys.\/} {\bf 21}, 41 (1971).
\bibitem{dehghanietal}
A. Dehghani, H. Fakhri and  B. Mojaveri: \emph{ The minimum-uncertainty coherent states for Landau levels,  }
{\em J. Math. Phys.\/} {\bf  53}, 123527 (2012); \\
A. Dehghani and  B. Mojaveri: \emph{New physics in Landau levels}, {\em J.  Phys.  A: Math. Theor.\/} {\bf  46}, 385303 (2013).
\bibitem{bergeron}  H. Bergeron, J. P. Gazeau and A. Youssef: 
\emph{Are the Weyl and coherent state descriptions physically equivalent?, } 
 {\em Phys. Lett.  A\/}    {\bf 377}, 598-605 (2013).
\bibitem{fock}
V. Fock: \emph{Bemerkung zur Quantelung des harmonischen Oszillators im Magnetfeld, } \emph{Z. Phys.\/} {\bf 47}, 446 (1928).
\bibitem{darwin}
C. G. Darwin:  
\emph{The diamagnetism of the free electron, }  
\emph{Proc. Camb. Phil. Soc.\/} {\bf 27}, 86 (1930).
\bibitem{johnson-lippman}
M. H. Johnson and B. A. Lippmann:  \emph{Motion in a constant magnetic field, } {\em Phys. Rev.\/} {\bf 76}, 828 (1949).
\bibitem{gilmore}
R. Gilmore: \emph{Lie Groups, Lie algebras, and Some of their Applications}, Wiley, New York  1974. 
\bibitem{magnus}
W. Magnus, F. Oberhettinger and R. P. Soni: \emph{Formulas and Theorems for the Special Functions of Mathematical Physics}, 
Springer-Verlag, New York  1966. 
\bibitem{watson}
G. N. Watson:  \emph{A Treatise on the Theory of Bessel Functions},  Cambridge University Press, Cambridge   1995. 
\bibitem{ali-antoine-gazeau}
S. T. Ali, J. P. Antoine  and J. P. Gazeau:
 {\em Coherent States, Wavelets and their Generalizations\/}, $2$nd ed. 2013,  
Springer-Verlag, New York  2000.  
\bibitem{popov}
D. Popov: 
\emph{Barut-Girardello coherent states of the pseudo-harmonic oscillator, } 
{\em J. Phys. A: Math. Gen.\/} {\bf 34},  1-14 (2001).
\bibitem{gradshtein}
I. S. Gradshtein: \emph{Table of integrals, series and products},
5th ed.,  Academic Press  1994. 
\bibitem{brif-aryeh}
C. Brif  and Y. Ben-Aryeh:  \emph{Subcoherent $p$-representation for non-classical photon states,} 
{\em Quantum Opt.\/} {\bf 6}, 391-6 (1994).
\bibitem{gazbook09}
J. P. Gazeau:   {\em Coherent States in Quantum Physics\/},  Wiley-VCH, Berlin  2009. 
\bibitem{agh}
I. Aremua, J. P. Gazeau and M. N. Hounkonnou:  
\emph{Action-angle coherent states for quantum systems with cylindric phase space, 
}    
{\em J. Phys. A: Math. Theor.\/} {\bf 45},  335302 (2012).
\bibitem{mandel2}
H. J. Kimble, M. Dagenais and L. Mandel: \emph{Photon Antibunching in Resonance Fluorescence, } 
{\em Phys. Rev. Lett.\/} {\bf 39}, 691 (1977).
\bibitem{mandel1}
L. Mandel: \emph{Sub-Poissonian photon
statistics in resonance fluorescence, } {\em Opt. Lett.\/} {\bf 4}, 205 (1979).
\bibitem{mandel3}
R. Short and L. Mandel:   \emph{Observation of sub-Poissonian photon statistics, } {\em Phys. Rev. Lett.\/} {\bf 51}, 384 (1983).
\bibitem{diedrich-walther}
F. Diedrich and H. Walther: \emph{Nonclassical Radiation of a Single Stored Ion,  } {\em Phys. Rev. Lett.\/} {\bf 58}, 203 (1987).
\bibitem{mandel-wolf}
L. Mandel and E. Wolf: {\em Optical coherence and quantum optics\/}, Cambridge
University Press, Cambridge   1995. 
\bibitem{bajer-miranowicz}
J. Bajer and  A. Miranowicz: \emph{Sub-Poissonian photon statistics of higher harmonics: quantum predictions via classical trajectories,} 
{\em J. Opt. B: Quantum Semiclass. Opt.\/} {\bf 2},  L10 (2000).
\bibitem{truessartetal}
F. Treussart, R. All\'{e}aume,  V. Le  Floc'h, L. T. Xiao, J.-M. Courty and J. F. Roch:  
 \emph{Direct Measurement of the Photon Statistics of a Triggered Single Photon Source, }
{\em Phys. Rev. Lett.\/} {\bf 89}, 093601 (2002).
\bibitem{lounis-moerner}
B. Lounis and W. E. Moerner: \emph{Single Photons on Demand from a Single Molecule at Room Temperature, }  
{\em Nature\/} {\bf 407}, 491 (2000).
\bibitem{lietal}
G. Li, T. C. Zhang, Y. Li and J. M.  Wang: \emph{Photon statistics of light fields based on single-photon-counting modules, }  
{\em Phys. Rev. A\/}  {\bf 71}, 023807 (2005).
\bibitem{zhangetal}
X.-Z. Zhang, Z.-H. Wang, H. Li, Q. Wu, B.-Q. Tang, F. Gao and J.-J. Xu: 
\emph{Characterization of Photon Statistical Properties with Normalized Mandel Parameter,}   
{\em Chin. Phys. Lett.\/}  {\bf 25}, 3976 (2008).
\bibitem{husimi}
K. Husimi: \emph{Some Formal Properties of the Density Matrix,}  
{\em Proc. Phys. Math. Soc. Jpn\/}  {\bf 22}, 264 (1940).
\bibitem{nikiforov}
A. Nikiforov  and  V.  Ouvarov: \emph{\'{E}l\'{e}ments De la Th\'{e}orie des Fonctions Speciales},  Mir, Moscow   1976. 
\bibitem{anderson-halliiwell}
A. Anderson and J. J. Halliwell: \emph{Information-theoretic measure of uncertainty due to quantum and thermal fluctuations, }
{\em  Phys. Rev. D\/}  {\bf 48}, 2753 (1993).
\bibitem{wehrl}
A. Wehrl: \emph{On the relation between classical and quantum entropy, }  {\em Rep. Math. Phys.\/} {\bf 16}, 353 (1979).
\bibitem{lieb}
E.H. Lieb:  \emph{Proof of an entropy conjecture of Wehrl, } {\em Commun. Math. Phys.\/}  {\bf 62}, 35 (1978).
\bibitem{pennini et al}
F. Pennini, A. Plastino and S. Curilef: \emph{Fisher information, Wehrl entropy, and Landau
Diamagnetism, }
{\em Phys. Rev. B\/}  {\bf 71}, 024420  (2005).







\end{thebibliography}
\end{document}